\documentclass[iop]{emulateapj}

\usepackage[centertags] {amsmath}
\usepackage{amsthm}

\newcommand{\G}         {\ensuremath{ \textrm{G} }}
\newcommand{\FMD}        {\ensuremath{ \textrm{FMD} }}
\newcommand{\I}         {{i}\,}
\newcommand{\kpc}       {\textrm{kpc}}

\newcommand{\dex}[1]    {\ensuremath{\times\textrm{10}^{#1}}}

\newcommand{\abs}[1]    {\ensuremath{\mid}{#1}\ensuremath{\mid}}
\newcommand{\sgn}       {\textrm{sgn}}

\newcommand{\arcsinh}     {\ensuremath{\textrm{arcsinh}}}
\newcommand{\arccot}      {\ensuremath{\textrm{arccot}}}

\newcommand{\li}  {\ensuremath{ \textrm{Li} }}

\newcommand {\HI}      {\textrm{H{\sc i}\ }}

\def\degK{\hbox{$^\circ\textrm{K}$}}

\usepackage[dvipsnames]{color}

\begin{document}

\title{The gravitational force and potential of the finite Mestel disk}
\author{Earl Schulz}
\affil{60 Mountain Road, North Granby, CT 06060} \email{earlschulz@gmail.com}

\begin{abstract}

Mestel determined the surface mass distribution of the finite disk for which the circular velocity is constant in the disk and found the
gravitational field for points in the $z=0$ plane. Here we find the exact closed form solutions for the potential and the gravitational
field of this disk in cylindrical coordinates over all the space.  The Finite Mestel Disk (FMD) is characterized by a cuspy mass
distribution in the inner disk region and by an exponential distribution in the outer region of the disk.  The FMD is quite different from
the better known exponential disk or the untruncated Mestel disk which, being infinite in extent, are not realistic models of real spiral
galaxies.  In particular, the FMD requires significantly less mass to explain a measured velocity curve.

\end{abstract}


\section{Introduction}

\cite{bin08} give a summary of some of the explicit potential-density pairs available for the study of spiral galaxies and the mathematical
methods used to discover these relationships. At the best of our knowledge, no finite thin-disk solutions are available, and this lack has
hampered the study of the dynamics of spiral galaxies. Solutions on the $z=0$ plane exist for the exponential disk and for the untruncated
Mestel disk \citep{bin08} and these have been most frequently used to model spiral galaxies. A major problem is that these disks of
infinite extent can over-predict the total mass and mass surface density associated with a measured velocity profile.

\cite{sch09} recently presented the complete closed form solution of the Maclaurin disk and a few related disks over all the space.  The
Maclaurin disk is applicable to the study of spiral galaxies for which the rotational velocity rises to the edge of the disk as is seen in
irregular galaxies.

Here we present the explicit expressions for the potential and gravitational field  of the finite Mestel disk (hereafter FMD) which was
first introduced by \cite{mes63}.  Until now the force vectors of the FMD were only known for points in the plane of the disk, and there
was no analytical solution of the potential.

Throughout this work we use the notation that $M$ is the total disk mass, $\alpha$ is the disk radius, $G$ is the gravitational constant,
and $(R,z)$ are the cylindrical co-ordinates in each meridional plane.

\section{The Solution of the Finite Mestel Disk}

\cite{sch09} performed  integral transformations of the mass distribution and potential of the \emph{collapsed homeoidal disk} with respect
to a parameter, determining explicit formula for the potential of a family of disks which includes the Maclaurin disk.  Here we perform a
similar integration of the collapsed homeoidal disk to find the potential and force vectors of the FMD over all the space.

\cite{mes63} constructed the mass distribution of the FMD which results in constant circular velocity over the entire disk. \cite{lyn78}
gives a convenient form for the surface mass density of the FMD:
\begin{equation} \label{eq:SurfDensFMD}
 \Sigma_{\FMD}(R)  =   \frac{M }{2 \pi \alpha R} \, \arccos\left(\frac{R}\alpha\right) ,~~~~~ R \le \alpha.
\end{equation}

Remarkably, the mass distribution given by equation (\ref{eq:SurfDensFMD}) can be obtained by an integral transformation of the surface
density with respect to $\alpha$ of the homeoidal disk:
\begin{equation} \label{eq:SurfDensHom2}
 \Sigma_{\hom}(R)  =   \frac{M }{2\pi \alpha^2 \sqrt{1-R^2/\alpha^2}} ,~~~~~   R \le  \alpha,
\end{equation}
i.e.:
\begin{equation}  \label{eq:TransformSigma} \Sigma_{\FMD}(R) = \frac{1}{\alpha} \int_0^\alpha{ \Sigma_{\hom}(R) d\alpha}.  \end{equation}
Since this is a linear transformation, the potential and gravitational fields of the FMD can be found by applying the same transformation
to the gravitational field and potential of the collapsed homeoidal disk:
\begin{equation}  \label{eq:TransformF} \pmb{F}_{\FMD}(R,z) = \frac{1}{\alpha} \int_0^\alpha{ \pmb{F}_{\hom}(R,z) d\alpha} , \end{equation}
\begin{equation}  \label{eq:TransformPhi} \Phi_{\FMD}(R,z) = \frac{1}{\alpha} \int_0^\alpha{ \Phi_{\hom}(R,z) d\alpha} . \end{equation}
The explicit expressions for $\pmb{F}_{\hom}$ and $\Phi_{\hom}$ are given in \cite{sch09}.

Integration of equation (\ref{eq:TransformF}) can be performed using elementary functions only. The expressions for the force vectors are
important because these equations can be used to describe the orbits in the gravitational field of the FMD.  The radial and vertical
components of the force are:
 \begin{equation}\label{eq:FRFMD-R}
    F_{R,\FMD}(R,z) =   -\frac{M \G}{R \alpha} \bigg[ \arcsin\left( T1 \right)     -\frac{\abs{z}}{ \sqrt{R^2+z^2} }   \arccos\left(  T2 \right)\bigg]
 \end{equation}
and
 \begin{equation}\label{eq:FzFMD-R}
    F_{z,\FMD}(R,z) =    - \frac{M \G~\sgn(z)}{\alpha \sqrt{R^2+z^2} }   \arccos\left(   T2   \right),
 \end{equation}
 where
   $$T1= \frac{ \sqrt{z^2+(\alpha+R)^2} - \sqrt{z^2+(\alpha-R)^2}  }{2 R} ,    $$
 and
   $$T2=     \frac{\sqrt{R^2+z^2}}{\alpha}   T1   . $$

In the z=0 plane equation (\ref{eq:FRFMD-R}) simplifies to:
\begin{equation}\begin{split}
 \label{eq:FRFMDonplane}
F_{R,\FMD}(R, 0) &= - \frac{\pi \G M } {2 \alpha R}        ,~~~~~~~~~~~~~~~~  R\leq\alpha , \\
                       &= - \frac{ \G M } {\alpha R } \arcsin\left( \frac{\alpha}{R}  \right) ,~~   R\geq\alpha,
\end{split}\end{equation}
which is identical except for sign to the formula given in \cite{mes63}. %equation 57
On the $z$  axis the equation (\ref{eq:FzFMD-R}) simplifies to:
\begin{align}
 \label{eq:FzFMDonaxis}
 F_{z,\FMD}( 0,z ) &=     -\frac{ \G M} { \alpha z}\arccot\left( \frac{\abs{z}}{\alpha}\right).
\end{align}

The evaluation of equation (\ref{eq:TransformPhi}) for the potential of the FMD is summarized in the Appendix. At variance with the force,
the potential (equation (\ref{eq:finalPhiFMD})), cannot be expressed using only elementary functions and involves the dilogarithm function
at complex arguments. We recall that the Euler dilogarithm function \citep{max03} is defined as
  \begin{equation} \label{eq:DilogDef}
       \li_2(\nu)=-\int^\nu_0{ \frac{\ln(1-t)}{t} dt}
  \end{equation}
and is a special case of the polylogarithm function defined by
 $\li_{s+1}(\nu)=\int^\nu_0\frac{\li_s(t)}{t}dt,$ with $ \li_1(\nu)=-\ln(1-\nu).$

\begin{figure*}[t]
  \plotone{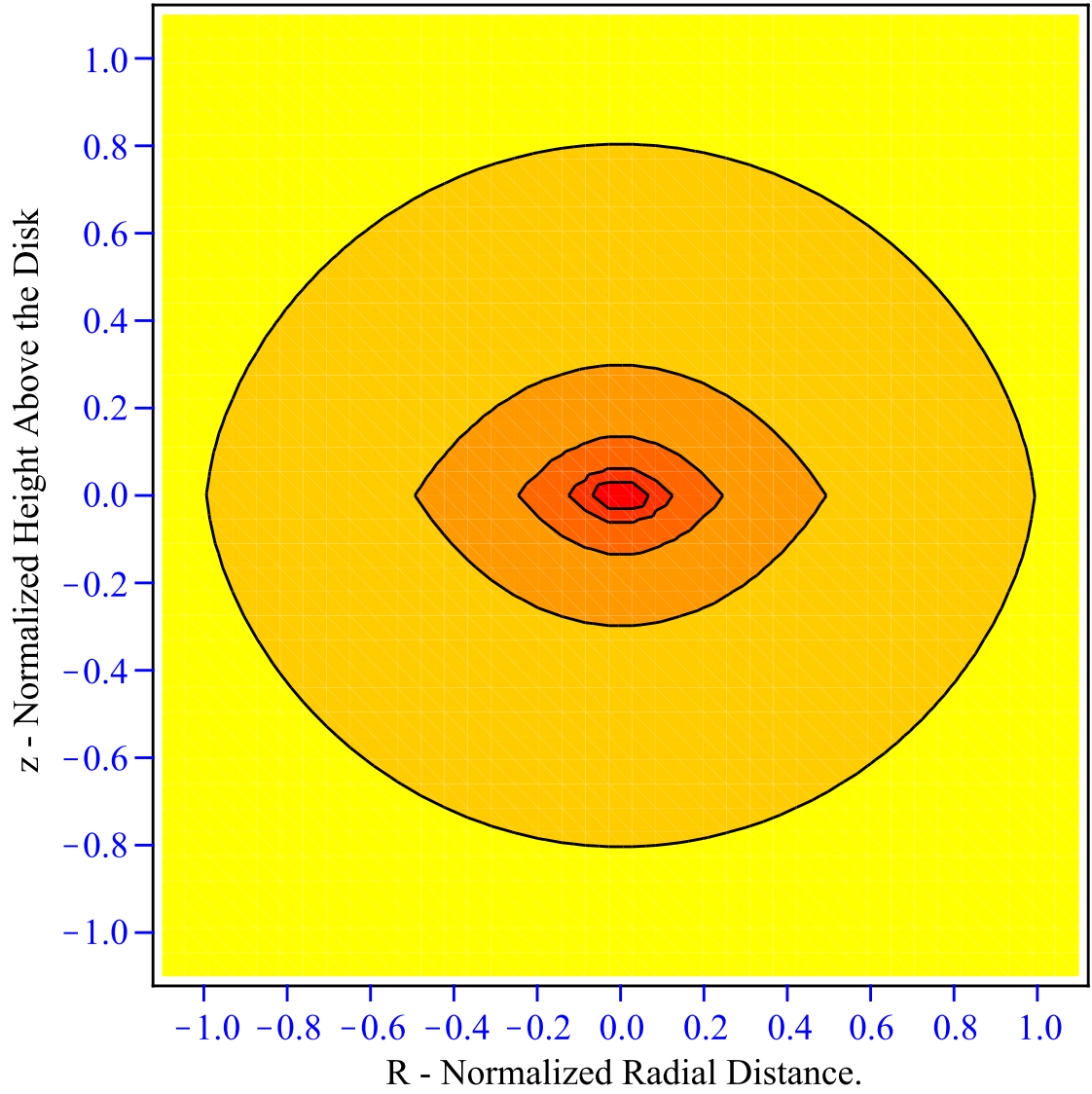}
  \caption{Potential of the finite Mestel disk.  The isopotential which passes through the edge of the disk is flattened by a factor of 0.81.  The isopotential
   through (0.2,0.0) is flattened by a factor of  0.53.  The potential is flattened near the disk but quickly becomes quite rounded.  }
  \label{fig:Potential}
\end{figure*}

 Figure \ref{fig:Potential} provides a contour plot of the FMD potential at several contour lines which are equally spaced in terms of potential
 difference.  The isopotential line which passes through the edge of the disk is flattened by a factor of about 0.81.

\subsection{Asymptotic Series Expansions}

The potential of the FMD is not analytic on the disk and so it is impossible to express the potential as a Taylor series expansion.
Instead, asymptotic expansions of the potential and force vectors were found in terms of power series of $1/R^n$ and $1/z^n$ for $R
\rightarrow \infty$ and for $z \rightarrow \infty$.

The first few terms of the series expansion of the radial force vector in the $z=0$ plane are:
\begin{equation}\label{eq:FR_series}
F_{R,\FMD}(R, 0) = -G M \left(  \frac{1}{R^2} + \frac{\alpha^2}{6R^4} + \frac{3\alpha^4}{40R^6}+\ldots \right).
\end{equation}

\begin{figure*}[t]
  \plotone{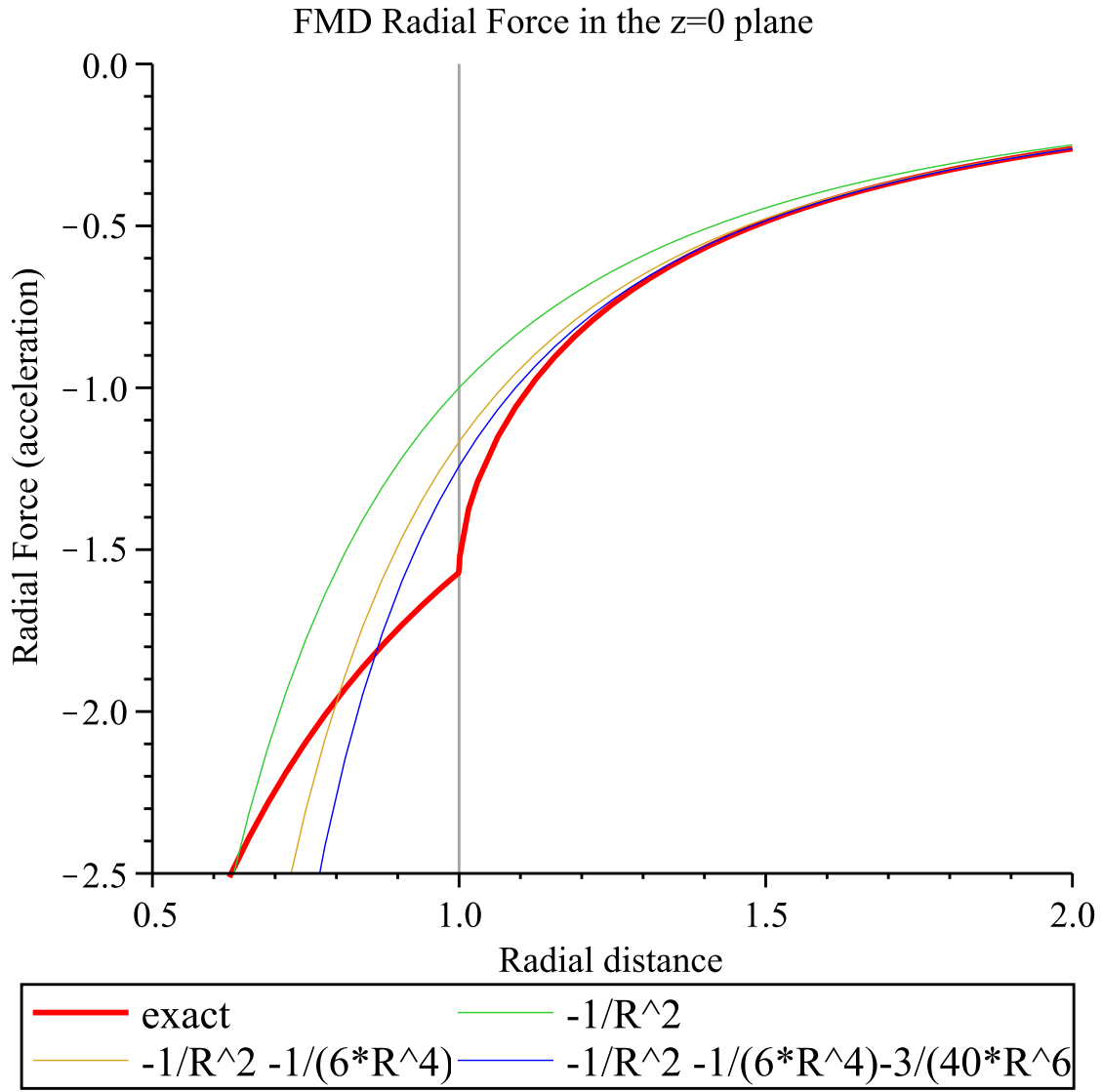}
  \caption{Radial Force Vector of the finite Mestel disk in the plane of the disk.  The radial force the disk edge $(\alpha=1)$ is greater than the Keplerian $1/R^2$ curve by a factor of $1.6$.  Beyond the
  disk, the force decreases quickly and joins the Keplerian curve by about $R=1.3\alpha$.  More terms of the series expansion improve the fit somewhat for points beyond the disk
  edge.}
  \label{fig:RadialForce}
\end{figure*}

Figure \ref{fig:RadialForce} shows the behavior of the radial force vector near the edge of the disk and also illustrates the convergence
properties of the asymptotic series. As expected, the radial force decreases more quickly than Keplerian beyond the edge of the disk and
quickly approaches the asymptotic behavior $F \propto 1/R^2$.  Including more terms increases the accuracy of the partial sum in the region
near the edge of the disk but convergence is slow and the series is not useful for computation.

The first few terms of the series expansion of the axial force vector on the $z$ axis are:
\begin{equation}\label{eq:Fz_series}
F_{z,\FMD}(0, z) = -G M \left(  \frac{1}{z^2} - \frac{\alpha^2}{3z^4} + \frac{\alpha^4}{5z^6}+\ldots \right).
\end{equation}

Unlike equation (\ref{eq:FR_series}), equation (\ref{eq:Fz_series}) is an alternating series and has somewhat different convergence
behavior.  The axial force decreases \emph{less} quickly than Keplerian beyond the limit of convergence at $z=\alpha$ to approach the
asymptotic behavior $F \propto 1/z^2$.

\subsection{Numerical Testing}

Numerical testing was performed in order to verify the analytical solution of the FMD. Normalized expressions for the potential and force
field were defined by choosing units such that the disk mass, the gravitational constant, and the disk radius were all 1.0. The
verification consisted of the following tests, all of which were performed with 80 digit precision:

\begin{list}{$\bullet$}{ }

\item

The Laplace equation,
 \begin{equation} \label{eq:laplace_eq}
 \frac{1}{R}\frac{\partial}{\partial R}\Phi(R,z)+\frac{\partial^2}{\partial R^2}\Phi(R,z)+\frac{\partial^2}{\partial z^2}\Phi(R,z) =0,
 \end{equation}
was verified over a $101\times101$ mesh in the interior of the region $(0 \le R \le 5,0 < z\le 5)$.   Equation (\ref{eq:laplace_eq}) was
evaluated numerically at each  of the  mesh points using standard finite difference approximations and using $\Phi$ as given by equation
(\ref{eq:finalPhiFMD}).  The derivative of the potential cannot be calculated for points  $R=0$ or $z=0$ and so these points were offset by
a small amount. The largest calculated value of the Laplace equation was $1.9\dex{-8}$ showing that equation (\ref{eq:laplace_eq}) is
satisfied over the entire space outside the disk.

\item

As shown in the Appendix, the potential of the FMD is a real-valued function of real variables which involves complex components.  The
imaginary parts cancel analytically but, due to roundoff, the calculated results might include an imaginary component.  It was verified
that the imaginary components were negligible at all points.

\item

The force field is defined by the relationship \par{$ \pmb{F} =  -\pmb{\nabla} \Phi$}. This relation was evaluated  numerically and the
resulting values compared to the force vectors given by (\ref{eq:FRFMD-R}) and (\ref{eq:FzFMD-R}).  The evaluation was performed at each of
the points of the $101 \times 101$ mesh described above and the difference was negligibly small.

\item

From the Laplace equation for a flat disk
 \begin{equation} \label{eq:limit_Fz}
 \lim_{z\to 0+} F_z(R,z) = -2 \pi G \Sigma(R).  % See \cite{bin08}  problem 2.3.
 \end{equation}
This relation was checked numerically at 100 radial points and the difference was negligibly small.

\end{list}

These numerical tests confirm that the analytical expressions for the solution of the FMD presented here are correct.

\section{Discussion}

A fundamental problem of the study of galaxy dynamics is to determine the mass model which best fits observational data.  The FMD is the an
appropriate model of a spiral galaxy in that both the mass distribution and the velocity curve of this disk match those of a typical spiral
galaxy.

\begin{figure*}[t]
  \plotone{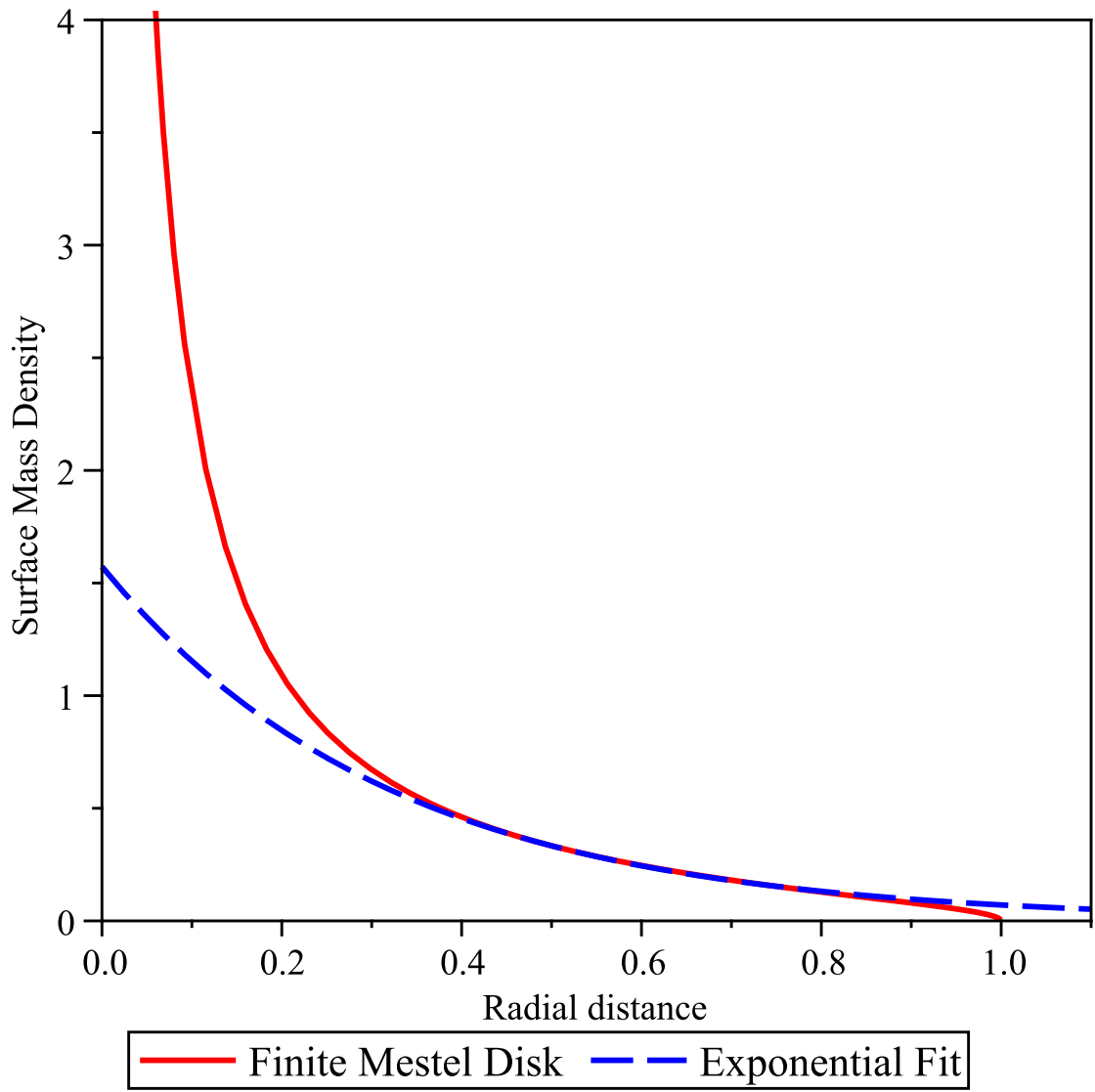}
  \caption{Normalized surface mass distribution of the finite Mestel disk.  The circular velocity of this disk is constant for $0\ge R\le \alpha$.  The dashed line is an exponential which was
  fit analytically at $R=\alpha/2$.  The curves are identical in the disk region beyond $R \geq0.2 \alpha$}.
  \label{fig:MassDist}
\end{figure*}

\begin{list}{$\bullet$}{ }
\item The flat velocity curve of the FMD is typical of spiral galaxies.

Although there is a lot of variation in galaxy rotation curves, most are flat or nearly so in the outer parts of the stellar disk.
 \cite{sof96} provides high-resolution position-velocity curves of nearby spiral galaxies based on CO line emission for the inner few \kpc~
and on \HI observations for the outer region. They show that, in general, the rotation curves of spiral galaxies are nearly flat, even in
the innermost region where the curve rises steeply within a few hundred parsecs.

The flat rotation curve of spiral galaxies might be the expected result of evolution.  \cite{sai90}  demonstrates that a flat rotation
curve and exponential  mass distribution results from the viscous contraction of a protogalactic disk with ongoing star formation. Working
in the context of an infinite disk, \cite{ber99}, finds that viscous  flows in a self-gravitating disk and ongoing star formation result in
a flat rotation curve.

\item
 The mass distribution of the FMD approximates the observed mass distribution of spiral galaxies.

The mass distribution of spiral galaxies generally follow Freeman's law \citep{fre70} that luminosity profiles of disk galaxies   usually
decrease  exponentially beyond the inner region: $\Sigma(R) \sim \exp^{-R/h_0}$.  Very often this exponential disk surrounds a central
bulge and a massive central black hole.  The FMD is a truncated exponential disk with a concentrated central region. Figure
\ref{fig:MassDist} plots the normalized surface mass distribution of the FMD given by equation (\ref{eq:SurfDensFMD}). Beyond about $ 0.2\,
\alpha$ the distribution approaches $\Sigma \propto \exp(-R/h_0)$ so closely that the two functions are indistinguishable.

The total mass of the FMD is:
\begin{equation} \label{eq:MassTotFMD}
M  = \int^\alpha_0{ 2 \pi R \Sigma_{\FMD}(R) d R} = \frac{ 2 \alpha {V_c}^2}{ \pi G}
\end{equation}
Where $\alpha$ is the radius of the disk and $V_c$ is the circular velocity which is constant over the disk.  In comparison, an assumed
spheroidal mass distribution requires a greater mass, by factor of $\pi/2\sim1.6$, to explain the same circular velocity profile.

The scale length of the exponential fit of the surface mass distribution shown in figure \ref{fig:MassDist}  is about one third of the
radius of the disk: $h_0=\alpha/3.1$. This applies to the outer region of the disk where the surface mass distribution of the FMD is
indistinguishable from that of an exponential disk.  In principle, this ratio could be confirmed by observation, noting that  the factor
$1/3.1$  applies to the scale length of the  \textbf{mass} distribution not the scale length of the luminosity function.

29\% of the total mass of the FMD is within the inner region $R\le0.2\alpha$ where the distribution approaches $\Sigma \propto 1/R$.
Identifying this region with the bulge of a spiral galaxy, it follows that the rotational velocity can be used to calculate directly both
the mass of the bulge and the mass of the disk.  This is significant because these two regions typically have different mass-to-light
ratios which are difficult to constrain.
\end{list}

The FMD is a better representation of a spiral galaxy than the simple exponential disk with $\Sigma \sim e^{-R/h_0}$ because the
exponential disk in infinite in extent and does not have a flat rotation curve.  Of course, many studies have used the so-called "maximum
disk" model consisting of an exponential disk, a bulge component and a dark halo component which can reproduce almost any rotation curve.
However, the procedure of fitting a rotation curve using these several components is under-constrained.  The masses of the disk, halo, and
bulge are free parameters as are the disk scale length and the several free parameters which describe the mass distributions of the bulge
and halo. More than one solution is possible, depending on the analyst's choice of the free parameters.

The FMD is often conflated with the more commonly used untruncated Mestel disk which was also introduced in \cite{mes63}.  However the two
disks are quite different.  The untruncated Mestel disk has a constant circular velocity but has infinite mass, is infinite in extent, and
the mass distribution $\Sigma=\sigma_0/R$ does not approximate the distribution seen for physical disks.  Despite these limitations, the
untruncated Mestel disk has the appeal of simplicity and is often chosen for various studies.

A flat disk model requires less mass than a spheroidal mass distribution to explain the same rotation curve. Table \ref{tab:GalMass}
compares three disk models which can be used to model disk galaxies and shows that a factor of $\sim0.64$, is required to correct the
result of a spherical model to that of a FMD for the case of a flat rotation curve. In the case of a rising velocity curve, a correction of
$\sim 0.42$ is required to correct the result a spherical model to that of a Maclaurin disk. These large errors which can arise from an
improper choice of a disk model are often neglected but should be kept in mind when interpreting rotational velocity data.

The Maclaurin disk is one of the few finite disks which have been completely solved \citep{sch09} and which, along with the FMD, is a
reasonable model of a spiral galaxy.  The Maclaurin disk, also known as the Kalnajs disk \citep{kal72} results from the collapse of a
rotating spheroid onto the plane.  The Maclaurin spheroid has been studied since the time of Newton \citep[e.g.,][for a historical summary
and discussions of the use of the Maclaurin spheroid and disk to study the stability of rotating bodies]{cha87,bin08,ber00}.

The  Maclaurin disk is defined by
\begin{equation} \label{eq:MassTotMac}
\Sigma = \frac{3 M}{2 \pi \alpha^2} \sqrt{1 - {R^2}/{\alpha^2}}
\end{equation}

This distribution is flat at the origin and falls off relatively slowly toward the edge of the disk.  The circular velocity rises linearly
to a maximum at the edge of the disk. This flat central surface density and linearly rising velocity curve is characteristic of irregular
galaxies. The total mass of the Maclaurin disk can be found from the velocity at the edge of the disk with the  formula
 $4 \alpha V_e^2/(3 \pi G) $.
In comparison an assumed spheroidal mass distribution would require a greater mass, by factor of $ 3 \pi / 4   \sim  2.4$, to explain the
same circular velocity profile. The large factor which must be applied in the case of a rising velocity curve has been neglected and has
lead to the prediction of large amounts of dark matter in irregular galaxies.
% See for instance \cite{bou07} which neglects this factor and claims that dark matter dominates the mass distribution of tidal debris dwarf galaxies by a factor of 2 to 3.

Dark matter halos are not required to explain the rotation curves of the stellar disks of disk galaxies \citep{pal00,ber00,cre98}. The best
evidence for dark matter halos comes from the velocity curves of the \HI gas which extends  beyond the stellar disks. This gas is often
seen to have a flat rotation curve far beyond the distance at which the gravitational attraction of the disk and bulge should be Keplerian.
There are a few alternate explanations of the velocity of extended \HI disks.  First, the MOND (Modified Newtonian Dynamics) hypothesis of
\cite{mil83} proposes that Newtonian mechanics break down below a threshold acceleration $a_0$.  Beyond this threshold the acceleration due
to gravity is proportional to $1/R$ rather than $1/R^2$.  Second, the Magnetic Hypothesis \citep{nel88,bat00,bat02} proposes that the
rotation of the \HI gas in the outer part of disk galaxies is governed by magnetism rather than gravity.  This subject is not yet
understood very well and FMD model presented may help decide among the several the several alternative explanations of the kinematics of
the outer \HI gas.

\subsection{Applications}

The FMD is a suitable model of the galaxy potential and force field of typical spiral galaxy and it is expected that the solution presented
here can be applied to several current astrophysical problems.  Some current areas of study which would benefit from a more physically
meaningful model of the gravitational potential of a disk include:

\begin{list}{$\bullet$}{}
\item The ``galactic fountain" model of gas flow within a disk galaxy \citep{sha76,bre80::1,bre80::2} holds that hot gas is ejected
    from
    the galactic disk by supernova explosions and part of this gas condenses into clouds which fall back onto the disk at high velocity.
    This model is widely accepted and some progress has been made to develop more details of the transport of
    gas and metals and the resulting chemical evolution of the disk.  See especially \cite{spi08} and \cite{spi11} and references therein.

The rotational velocity of extra-planar gas in spiral galaxies is observed to decrease linearly with height above the
    disk\citep{fra05,ran97,ran05} and this behavior has not yet been understood completely.  \cite{bar06} compares two models of
    the extra-planar gas which attempt to describe the decrease of velocity with height: a ballistic model and a baroclynic model.
    The ballistic model did not reproduce the data while the baroclynic model was more successful.  More recently, \cite{Mar11}
    modeled with partial success the case in which cold $(\sim10^4 \, \degK)$ gas is ejected from the disk and interacts with hot
    $(\sim10^6 \, \degK)$ gas in the halo.

    These studies would benefit from the application of the FMD model which predicts that the radial
    attraction of a disk-dominated galaxy decreases with increasing height in a way which explains most of the observed effect.

\item  \cite{mal08} and \cite{dej88} investigate orbits in the potential of a galaxy disk especially as applicable to the study  globular
    clusters. The potential of the disk is assumed to be the \cite{kuz56} potential
    $$ \Phi(R,z) = -G M \left[ R^2+  ( \abs{z}+z_0)^2 \right]^{-\onehalf} $$
    and the main parameters of the disk were determined based on a fit to the measured data of \cite{dau96}.  A galaxy model based on the
    FMD would provide more relevant information for this sort of study, especially near the edge of the stellar disk.
\end{list}

\section{Conclusions}

We have presented the complete closed form solution in cylindrical coordinates for the gravitational potential and force field of the FMD.
The mass distribution and the rotation curve of the FMD match those of typical spiral galaxies and so this work provide a useful model for
the study of the properties of disk galaxies.

The FMD is defined by two parameters, disk mass and disk diameter, in contrast to the commonly used  maximum disc solution which permits
many free parameters.  This simpler description will allow us to standardize the characterization of populations of spiral galaxies which
will aid in the study of, e.g., relations such as the baryonic Tully-Fisher relation or the Magnitude-Diameter relation.

\acknowledgments
Maple 11 was used to help solve the complicated integrals which were encountered and to perform the high precision numeric
verification.

\newpage

\begin{appendix}

%%\textbf{\red{ Copy Editor!! Please set this appendix one-column} }

\section{The Potential of the Finite Mestel Disk}\label{app:Derivation}

We start reporting the potential for the collapsed homeoidal disk \citep{sch09}:
\begin{equation}  \label{eq:PhiHom}
   \Phi_{\hom}(R,z) =  -\frac{G M}{2 \alpha} \left[\arcsin\left(\frac{ \alpha-\I  z}{R}\right)+\arcsin\left(\frac{\alpha+\I  z}{R}\right) \right],
\end{equation}
where the disk surface density is given in equation (\ref{eq:SurfDensHom2}).

From the linearity of equation (\ref{eq:TransformSigma}), it follows that the potential of the finite Mestel disk is found by applying the
transformation (\ref{eq:TransformSigma}) to the potential of the collapsed homeoidal disk;
\begin{align}
 \label{eq:PhiFMD0} \Phi_{\FMD}(R,z) & = \frac{1}{ \alpha}\int^ \alpha_{0}{\Phi_{\hom}(R,z)  d\alpha}    ,\\
 \label{eq:PhiFMD1}          & = \frac{-G M}{2\alpha} \int^\alpha_0 \frac{\arcsin\left(\frac{\alpha+\I  z}{R} \right) + \arcsin\left(\frac{\alpha-\I z}{R} \right)}{ \alpha} d\alpha .
\end{align}

$\Phi_{\FMD}$ is a well-behaved function of the parameter $\alpha$ for arbitrary  non-zero $R$ and $z$.  That is, the potential at a
spatial point (R,z) not on the disk changes smoothly as the disk radius is varied.  In fact, $\Phi_{\FMD}$ is an \emph{analytic} function
of $\alpha$ for complex $\alpha$ over the entire complex plane except for points on the disk.

We construct the solution to (\ref{eq:PhiFMD1}) using the solution of the simpler integral
\begin{equation}\label{eq:phi1-def}
\phi_1(\beta,\zeta)  =   \int^{\beta}_\epsilon  \frac{\arcsin\left( t+\I  \zeta \right)  }{t} dt   ,
\end{equation}
where the integration path for equation (\ref{eq:phi1-def}) is along a line segment on the real axis.  Here the parameter $\epsilon$ was
introduced to avoid the pole of the integrand  at $t = 0$ and we take $\zeta \neq 0$, $0< \epsilon < \beta < 1$, to ensure that the
integrand is analytic in a region surrounding the path.

The solution to equation (\ref{eq:PhiFMD1}) is then:
\begin{equation}\label{eq:PhiFMD-sol}
 \Phi_{\FMD}(R,z) = -\frac{G M}{2 \alpha}~ \lim_{\epsilon \to 0}{   \left[ \phi_1(\alpha/R,z/R)+\phi_1(\alpha/R,-z/R) \right] }.
\end{equation}

Make the change of variable $w=\arcsin(t +\I  \zeta) $:
\begin{equation}\label{eq:phi1-w}
 \phi_1(\beta,\zeta)   = \int^{w_1}_{w_0}{ \frac{ w \cos(w) }{\sin(w) -\I  \zeta}}dw  ,
\end{equation}
where $w_1 =\arcsin(\alpha+\I  \zeta)$ and $w_0 =\arcsin(\epsilon+\I  \zeta)$. The path of integration of equation (\ref{eq:phi1-w}) is
taken the straight line segment connecting points $w_1$ and $w_0$. The transformation is one-to-one near the integration path and so this
is a \emph{conformal mapping} from one analytic domain to another.

Expand the integrand of (\ref{eq:phi1-w}) using the identities $\cos(w) = ( e^{\I  w}  +e^{-\I  w})/2    $ and  $\sin(w) = ( e^{\I  w}
-e^{-\I  w})/2   $:
 \begin{align}
\phi_1(\beta,\zeta)    &=  \int^{w_1}_{w_0}{ \frac{\I w (e^{2\I  w} +1)}{e^{2\I  w} +2 \zeta e^{\I  w} -1} dw}             .\\
 \intertext{Define $\lambda=\sqrt{\zeta^2+1}+\zeta $ so that $1/\lambda=\sqrt{\zeta^2+1}-\zeta$:}
 \phi_1(\beta,\zeta)   &=  \int^{w_1}_{w_0}{ \frac{\I  w (e^{2\I  w} +1)}{ (e^{\I  w}+\lambda) (e^{\I  w}-1/\lambda) } dw}  .\\
 \intertext{Expand the integrand: }
 \phi_1(\beta,\zeta)    &= \frac{\I  \lambda}{1+\lambda^2}\int^{w_1}_{w_0}{ \left( - \frac{w}{e^{\I  w}+\lambda} +\frac{w}{e^{\I  w} -1/\lambda}  -\frac{w e^{2\I  w}}{e^{\I  w} +\lambda}  +\frac{w e^{2\I  w}}{e^{\I  w} -1/\lambda}   \right) dw } .
\end{align}

The integral can be solved by using the formulae:
\begin{align}
 \label{eq:li2int1}  \int{\frac{t}{e^{t}+c} dt} &= \frac{t^2}{ 2 c} -   \frac{t \ln(1+  {e^{t}}/{c})}{c}     -\frac{\li_2\left( - {e^{t}}/{c} \right) }{c} ,\\
 \label{eq:li2int2}  \int{\frac{t e^{2 t}}{e^{t} + c} dt} &=  (t-1) e^t -  {c t \ln(1+  {e^{t}}/{c})}      -c\li_2\left( - {e^{t}}/{c} \right).
\end{align}

Equations (\ref{eq:li2int1}) and (\ref{eq:li2int2}) can be proved by differentiation using the definition of $\li_2$, the dilogarithm
function, given by equation (\ref{eq:DilogDef}).

Collect terms:
\begin{equation}
 \phi_1(\beta,\zeta) =   -\I \li_2(e^{\I w}\lambda)  -\I \li_2(-e^{\I w}/\lambda) -\I w^2/2 +w\ln(1+e^{\I w}/\lambda) +w\ln(1-\lambda e^{\I w} ) \Bigg|^{w_1}_{w_0}  .
\end{equation}

Evaluate this expression, make the substitution $\lambda=\sqrt{1+\zeta^2} + \zeta$, and simplify the result using the identity
\citep[1.622]{gra07}: %
\begin{equation}
  \I \ln(t-\I \sqrt{1-t^2}) = \arccos(t).
\end{equation}

The solution to equation (\ref{eq:phi1-w}) is :
\begin{equation}\label{eq:finalphi1}
       \phi_1(\beta,\zeta) =  \phi_2(\beta,\zeta) -\phi_2(\epsilon,\zeta),
\end{equation}

where
  \begin{equation}\label{eq:PhiFMDdd}\begin{split}
  \phi_2(\beta, \zeta) =
     & -\I\li_2\left\{ { \Big[ \sqrt{1 +(\I\beta+\zeta)^2} -\zeta+\I\beta \Big]  \Big(\zeta-\sqrt{1+\zeta^2} \Big) }  \right\} \\
     & -\I\li_2\left\{ { \Big[ \sqrt{1 +(\I\beta+\zeta)^2} -\zeta+\I\beta \Big]  \Big(\zeta+\sqrt{1+\zeta^2} \Big) }  \right\} \\
     &  +\frac{1}{2} \I\arcsin^2\left( {\beta+\I \zeta} \right)
        +\left[ \ln\left( {2\beta} \right) -  { \pi}{2}\I\right]  \arcsin\left( {\beta+\I \zeta} \right)  .
\end{split}\end{equation}

The limit  at $\epsilon \to 0$ is required.  After some simplifications we find that:
 \begin{equation} \label{eq:epsilonLimit}
  \lim_{\epsilon \to 0} [\phi_2(\epsilon,\zeta) + \phi_2(\epsilon,-\zeta)  =  \I\arcsinh^2(\zeta) -\frac16\pi^2\ .
  \end{equation}

Use equations (\ref{eq:finalphi1}) and (\ref{eq:epsilonLimit}) to evaluate equation (\ref{eq:PhiFMD-sol}):
\begin{equation}\label{eq:finalPhiFMD}
        \Phi_{\FMD}(R,z) = \frac{-G M}{2\alpha} \left[   \phi_3( R, z) + \phi_3(R, -z)  -\I\arcsinh^2\left(\frac{z}{R}\right)  +\frac16\pi^2 \right]  , \end{equation}

where
  \begin{equation}\label{eq:phi3}\begin{split}
  \phi_3( R, z) =
     & -\I\li_2\left\{\frac{ \Big[ \sqrt{R^2 +(\I\alpha+z)^2} -z+\I\alpha \Big]  \Big(z-\sqrt{R^2+z^2} \Big) }{R^2} \right\} \\
     & -\I\li_2\left\{\frac{ \Big[ \sqrt{R^2 +(\I\alpha+z)^2} -z+\I\alpha \Big]  \Big(z+\sqrt{R^2+z^2} \Big) }{R^2} \right\} \\
     &  +\frac{1}{2} \I\arcsin^2\left(\frac{\alpha+\I z}{R}\right)
        +\left[ \ln\left( \frac{2\alpha}{R}\right) - \frac{ \pi}{2}\I\right]  \arcsin\left(\frac{\alpha+\I z}{R}\right)  .
\end{split}\end{equation}

The terms of equation (\ref{eq:finalPhiFMD}) become large near the $z=0$ plane and near the $R=0$ axis so that equation
(\ref{eq:finalPhiFMD}) is unusable those regions.  The potential in these regions was found by finding the limits of equation
(\ref{eq:finalPhiFMD}).

The potential in the $z=0$ plane is:
\begin{equation} \begin{split}\label{eq:PhiFMDonplane}
   \Phi_{\FMD}(R, 0) = &
  \frac{\pi G M}{2\alpha} \ln\left( \frac{R}{2\alpha}  \right),  \qquad R\leq \alpha, \\
  \Phi_{\FMD}(R, 0) = &
  -\frac{ G M}{\alpha}  \Bigg[
     \frac{\I \pi^2 }{12}
   + \frac{\I}{2} \arcsin^2\left( \frac { \alpha}{R} \right)
   + \ln\left(  \frac{2 \alpha}{R}  \right) \arcsin \left( {\frac {\alpha}{R}} \right)
   - \frac{\I\pi}{2} \, \arcsin \left( {\frac {\alpha}{R}} \right) \\&
   - \I\li_2 \left( -{ \frac {\sqrt{{R}^{2}-{\alpha}^{2}}+  \I\alpha}{R}} \right)
   - \I\li_2 \left(  { \frac {\sqrt{{R}^{2}-{\alpha}^{2}}+  \I\alpha}{R}} \right)
  \Bigg],  \qquad  R\geq \alpha.
\end{split} \end{equation}
The potential on the $z$ axis  is:
\begin{equation}
 \label{eq:PhiFMDonaxis}
   \Phi_{\FMD}( 0,z) = \frac{\I G M }{2 \alpha}\left[  \li_2\left( \frac{\I\alpha}{\abs{z}}\right) -\li_2\left( -\frac{\I\alpha}{\abs{z}} \right)   \right     ]  .
\end{equation}

\newpage

\end{appendix}

%%%%%%%%%%%%%%%%%%%%%%%%%%%%%%%%%%%%%%%%%%%%%%%%%%%%%%%
%%%%%%%%%%%%%%%%%%%%%%%%%%%%%%%%%%%%%%%%%%%%%%%%%%%%%%%

\clearpage

\clearpage

\newcommand{\D}{\displaystyle}

\newpage
\renewcommand{\arraystretch}{3.0}
{\begin{table} \label{tab:GalMass}
 \caption{\bfseries Total Mass of Disk Galaxies}\vspace{5pt}
\begin{tabular}{| l | c |c |l|}
\hline
\bfseries ~~  &\bfseries  Mass Distribution &\bfseries Total Disk Mass&\bfseries Velocity Curve\\
\hline
Central Spheroid     &  - - -                                                                &$ \D  \frac{\alpha V_e^2}{G}                 $&Falls as $1/\sqrt{R}$\\
Finite Mestel Disk   &$ \D \frac{M }{2 \pi \alpha R} \, \arccos\left(\frac{R}\alpha\right)  $&$ \D  \frac{2}{  \pi}\frac{\alpha V_e^2}{G}  $&Constant\\
Maclaurin disk       &$ \D \frac{3 M}{2 \pi \alpha^2} \sqrt{1 - {R^2}/{\alpha^2}}           $&$ \D  \frac{4}{3 \pi}\frac{\alpha V_e^2}{G}  $&Rises Linearly\\
\hline
\end{tabular}
\end{table}
}

\end{document}